\documentclass[a4paper,12pt]{article}
\usepackage[pctex32]{graphics}
\usepackage{amssymb,amsmath}

\begin{document}

\topmargin 0pt
\oddsidemargin 0mm
\newcommand{\be}{\begin{equation}}
\newcommand{\ee}{\end{equation}}
\newcommand{\ba}{\begin{eqnarray}}
\newcommand{\ea}{\end{eqnarray}}
\newcommand{\fr}{\frac}

\renewcommand{\thefootnote}{\fnsymbol{footnote}}

\begin{titlepage}

\vspace{5mm}
\begin{center}
{\Large \bf Logarithmic conformal field theory approach to
topologically massive gravity}

\vspace{12mm}

{\large   Yun Soo Myung\footnote{e-mail
 address: ysmyung@inje.ac.kr}}
 \\
\vspace{10mm}

{\em Institute of Basic Science and School of Computer Aided
Science,
\\Inje University, Gimhae 621-749, Korea}
\end{center}

\vspace{5mm}

\centerline{{\bf{Abstract}}}

\vspace{5mm}

We study the topologically massive gravity at the chiral point
(chiral gravity) by using the logarithmic conformal field theory.
Two new tensor fields of $\psi^{new}$ and $X$ are introduced for a
candidate of propagating physical field at the chiral point.
However, we show that ($\psi^{new},\psi^L$) form a dipole ghost
pair of unphysical fields and $X$ is not a primary. This implies
that there is no physically propagating degrees of freedom at the
chiral point.

\vspace{5mm}

\end{titlepage}

\newpage

\renewcommand{\thefootnote}{\arabic{footnote}}
\setcounter{footnote}{0} \setcounter{page}{2}

\section{Introduction}
The gravitational Chern-Simons terms in three-dimensional Einstein
gravity produces a physically propagating massive
graviton~\cite{DJT}. This theory with a negative cosmological
constant $\Lambda=-1/l^2$ gives us the AdS$_3$ solution. At the
chiral point of $\mu l=1$, that is, at the chiral gravity,  the
massive graviton becomes a massless left-moving graviton, an
unphysical propagation~\cite{LSS1,LSS2}.  However, several
counterexamples have been
reported~\cite{CDWW1,GJ,GKP,Park,GJJ,CDWW2,Carl}. Especially, we
wish to mention  two cases. Firstly,
 the non-chiral solution $\psi^{new}_{\mu\nu}=\partial_{\mu l}\psi^M_{\mu\nu}|_{\mu l=1}$
   was found  in global coordinates~\cite{GJ}
but it turned out to not  satisfy  the asymptotic  boundary
condition for chiral gravity~\cite{Stro}. Secondly, a related
family of $X_{\mu\nu}=\bar{L}_{-1}\psi^{new}_{\mu\nu}$ does satisfy the
asymptotic boundary condition~\cite{GKP} but it acts as a trivial gauge
degree of freedom~\cite{Stro}.

In this work, we address this issue again because  a key point in
the  chiral gravity is to find physically propagating  degrees of
freedom on the AdS$_3$ spacetime background. We show that there is
no physically propagating degrees of freedom  at the chiral point
by using the logarithmic conformal field theory (LCFT). Our result
supports that the original work of chiral gravity~\cite{LSS1} is
correct.

\section{Topologically massive gravity}
We start with the topologically massive gravity in anti-de Sitter
spacetimes (TMG)~\cite{DJT}
\begin{equation}
I_{\rm TMG}=\frac{1}{16 \pi G_3}\int
d^3x\sqrt{-g}\Bigg[R+\frac{2}{l^2}-\frac{1}{2\mu}\varepsilon^{\lambda\mu\nu}\Gamma^\rho_{~\lambda\sigma}
\Big(\partial_{\mu}\Gamma^\sigma_{~\nu\rho}+\frac{2}{3}\Gamma^\sigma_{~\mu\tau}\Gamma^\tau_{~\nu\rho}\Big)\Bigg],
\end{equation}
where $\varepsilon$ is the tensor density defined by
$\epsilon/\sqrt{-g}$ with $\epsilon^{012}=1$. The $1/\mu$-term is
the first higher derivative correction in three dimensions because
it is the third-order derivative.

Varying  the this action  leads to the Einstein equation
\begin{equation} \label{eineq}
G_{\mu\nu}+\frac{1}{\mu}C_{\mu\nu}=0,
\end{equation}
where the Einstein tensor including the cosmological constant is
given by
\begin{equation}
G_{\mu\nu}=R_{\mu\nu}-\frac{R}{2}g_{\mu\nu}
-\frac{1}{l^2}g_{\mu\nu}
\end{equation}
and the Cotton tensor is
\begin{equation}
C_{\mu\nu}= \varepsilon_\mu^{~\alpha\beta}\nabla_\alpha
\Big(R_{\beta\nu}-\frac{1}{4}g_{\beta\nu}R\Big).
\end{equation}
The  BTZ black hole solution~\cite{BTZ}  is given by
\begin{equation}
ds^2_{BTZ} = -f(r) dt^2 + \frac{dr^2}{f(r)} + r^2 \Big(d\phi +
N^\phi(r) dt\Big)^2 , \label{btzmetric}
\end{equation}
where the metric function $f$ and the lapse function $N^\phi$ are
\begin{equation}
f(r) = -8 G_3 m + \frac{r^2}{l^2} + \frac{16 G_3^2 j^2}{r^2},~~
N^\phi(r) = - \frac{4 G_3 J}{r^2}.\label{def_N}
\end{equation}
 Here $m$ and $j$ are the mass and angular momentum of the BTZ
 black hole, respectively. In this work, we consider the AdS$_3$ solution which appears  for
 $m=-1/8G_3,~j=0$ only. We note that the Cotton tensor $C_{\mu\nu}$ vanishes for
 any solution to Einstein gravity, so all solutions to general
 relativity are also solutions of TMG.

 However, TMG possesses physical degrees of freedom propagating
 on the AdS$_3$ spacetimes. In order to explore this, we study the
 perturbation around the AdS$_3$ spacetimes by considering the
 metric fluctuations
 \begin{equation}
 g_{\mu\nu}=\bar{g}_{\mu\nu}+h_{\mu\nu}
 \end{equation}
 where the AdS$_3$ metric appears as
\begin{equation}
ds^2_{AdS} =\bar{g}_{\mu\nu}dx^\mu dx^\nu=
-\Big(1+\frac{r^2}{l^2}\Big) dt^2 +
\frac{dr^2}{\Big(1+\frac{r^2}{l^2}\Big)} + r^2 d\phi^2.
\label{adsmetric}
\end{equation}
Introducing the global coordinates with $r=l\sinh\rho$ and $\tau=
lt$ as
\begin{equation}
ds^2_{global} =\bar{g}_{\mu\nu}dx^\mu dx^\nu=l^2\Big(-\cosh^2\rho
d\tau^2+\sinh^2\rho d\phi^2+d\rho^2\Big), \label{globalmetric}
\end{equation}
 then it covers the whole space with the boundary at
 $\rho=\infty$. The isometry group of AdS$_3$ space is
 $SL(2,R)\times \overline{SL(2,R)}$ and its generators are realized on
 scalar fields as
\begin{eqnarray}
L_0&=& i \partial_u,~~~L_{\pm1}=ie^{\pm
iu}\Big(\frac{\cosh2\rho}{\sinh2\rho}\partial_u-\frac{1}{\sinh2\rho}\partial_v
\mp\frac{i}{2}\partial_\rho\Big), \label{gene1} \\
\bar{L}_0&=& i \partial_v,~~~\bar{L}_{\pm1}=ie^{\pm
iv}\Big(\frac{\cosh2\rho}{\sinh2\rho}\partial_v-\frac{1}{\sinh2\rho}\partial_u
\mp\frac{i}{2}\partial_\rho\Big), \label{gene2}
\end{eqnarray}
where two light-cone coordinates $u/v=\tau\pm\phi$ are introduced
to study the boundary physics. Using the transverse and traceless
gauge $\bar{\nabla}_\alpha h^{\alpha\beta}=0$ with  $h=0$, the
linearized equation of ${\cal G}^{(1)}_{\mu\nu}+{\cal
C}^{(1)}_{\mu\nu}=0$ to the Einstein equation (\ref{eineq}) could
be converted into a compact form as the graviton equations of
motion~\cite{LSS1},
\begin{equation} \label{compeq}
\Big( {\cal D}^R {\cal D}^L {\cal D}^M h \Big)_{\mu\nu}=0
\end{equation}
where
\begin{equation}
\Big({\cal D}^{L/R}\big)_\mu~^\nu=\delta_\mu~^\nu \pm l
\varepsilon_{\mu}~^{\alpha \nu}\bar{\nabla}_\alpha,~~\Big({\cal
D}^{M}\Big)_\mu~^\nu=\delta_\mu~^\nu
+\frac{1}{\mu}\varepsilon_{\mu}~^{\alpha \nu}\bar{\nabla}_\alpha.
\end{equation}
Here the covariant derivative $\bar{\nabla}_\alpha$ is defined
with respect to $\bar{g}_{\mu\nu}$ in Eq.(\ref{globalmetric}). We
note that ${\cal D}^R, {\cal D}^L, {\cal D}^M$ commute with each
other. Then we could obtain all solutions of $h^{R/L}_{\mu\nu}$
and $h^{M}_{\mu\nu}$ which satisfy the first-order equations,
respectively,
\begin{equation} \label{firsteqs}
\Big( {\cal D}^R h^R\Big)_{\mu\nu}=0,~\Big( {\cal D}^L h^L
\Big)_{\mu\nu}=0,~\Big( {\cal D}^M h^M \Big)_{\mu\nu}=0,
\end{equation}
where the first two equations describe gauge degrees of freedom
(right-and left-movers), while the last one describes physically
propagating degrees of freedom (massive graviton).  However, at
the chiral point of $\mu l=1$, we find from ${\cal D}^M={\cal
D}^L$ that the massive graviton turns out to be the left-mover.
The linearized equation of ${\cal G}^{(1)}_{\mu\nu}+{\cal
C}^{(1)}_{\mu\nu}=0$ can be rewritten as
\begin{equation} \label{dlineq}
\Big({\cal D}^M {\cal G}^{(1)}\Big)_{\mu\nu}=0
\end{equation}
with \begin{equation} {\cal
G}^{(1)}_{\mu\nu}=-\frac{1}{2}\Big(\bar{\nabla}^2+\frac{2}{l^2}\Big)h_{\mu\nu}.
\end{equation}
Further, considering
\begin{equation}
\Big(\bar{{\cal D}}^M{\cal D}^M {\cal G}^{(1)}\Big)_{\mu\nu}=0
\end{equation}
with \begin{equation} \bar{{\cal D}}^M=\delta_\mu~^\nu
-\frac{1}{\mu}\varepsilon_{\mu}~^{\alpha \nu}\bar{\nabla}_\alpha
\end{equation}
leads to the fourth-order equation
\begin{equation} \label{fourthh}
\Big(\bar{\nabla}^2+\frac{3}{l^2}-\mu^2\Big)\Big(\bar{\nabla}^2+\frac{2}{l^2}\Big)h_{\mu\nu}=0.
\end{equation}
Eq.(\ref{dlineq}) could be derived from the bilinear action up to
total derivatives  as
\begin{eqnarray}
I_{\rm (2)}&=&-\frac{1}{32 \pi G_3}\int
d^3x\sqrt{-g}h^{\mu\nu}\Big( {\cal G}^{(1)}_{\mu\nu}+{\cal
C}^{(1)}_{\mu\nu}\Big) \nonumber \\
&=& \frac{1}{64 \pi G_3}\int d^3x\sqrt{-g}\Bigg[
-\bar{\nabla}^\lambda h^{\mu\nu} \bar{\nabla}_\lambda h_{\mu\nu}
-\frac{1}{\mu}\bar{\nabla}_\alpha
h^{\mu\nu}\epsilon_\mu^{~\alpha\beta}\Big(\bar{\nabla}^2+\frac{2}{l^2}\Big)h_{\beta\nu}
+ \frac{2}{l^2}h^{\mu\nu}h_{\mu\nu} \Bigg].
\end{eqnarray}
On the other hand, we can use  the first-order equations in
Eq.(\ref{firsteqs})  to have
\begin{equation}
\Big( {\cal D}^L{\cal D}^R h^R\Big)_{\mu\nu}=0,~\Big( {\cal
D}^R{\cal D}^L h^L \Big)_{\mu\nu}=0,~\Big( \bar{{\cal D}}^M{\cal
D}^M h^M \Big)_{\mu\nu}=0,
\end{equation}
which leads to the conventional second-order equations
\begin{equation} \label{sec-equation}
\Big(\bar{\nabla}^2+\frac{2}{l^2}\Big)h^{R/L}_{\mu\nu}=0,~
\Big(\bar{\nabla}^2+\frac{2}{l^2}+\frac{1}{l^2}-\mu^2\Big)h^{M}_{\mu\nu}=0.
\end{equation}
Here the first equation represents the massless gravitons of
right and left-movers, while the last one denotes the massive
graviton for $\mu^2 \not=1/l^2$. We note that
$\frac{2}{l^2}=-2\Lambda$ is not the mass term but it appears
because all gravitons are propagating on the AdS$_3$ spacetimes.
For the scalar propagation, there is no such a term. Also, these
second-order equations are insensitive to signs of $\pm l$ and $\pm
1/ \mu$, while the first-order equations are sensitive to
signs~\cite{ssol}.

 The explicit form of the wave function for a
massive primary graviton  could be  constructed by using
$SL(2,R)$-algebra. Considering
$\bar{\nabla}^2=-(2/l^2)(L^2+\bar{L}^2+3)$ acting on  a rank two
tensor, Eq.(\ref{sec-equation})
 leads to two algebraic equations  of weights $(h,\bar{h})$  for
primary fields
\begin{equation}
h(h-1)+\bar{h}(\bar{h}-1) -2=0,~~2h(h-1)+2\bar{h}(\bar{h}-1)
-3=\mu^2l^2.
\end{equation}
Solving these equations plus the gauge condition together with
asymptotic boundary condition, we have primary weights (2,0),
(0,2), and $(3/2+\mu l/2,-1/2+\mu l/2)$ with $h-\bar{h}=\pm 2$ for
left, right-moving massless gravitons, and massive graviton,
respectively. The massive primary states are constructed from
$L_1h_{\mu\nu}^M=\bar{L}_1h_{\mu\nu}^M=0$ as
\begin{equation}
h^M_{\mu\nu}= \Re \psi^M_{\mu\nu},
\end{equation}
where
\begin{equation}
 \psi^M_{\mu\nu}=e^{-(3/2+\mu l/2)iu-(-1/2+\mu
 l/2)iv}\frac{\sinh^2\rho}{(\cosh\rho)^{1+\mu l}}\left(%
\begin{array}{ccc}
  1 & 1 & ia \\
  1 & 1 & ia \\
  ia & ia & -a^2 \\
\end{array}%
\right)_{\mu\nu}
\end{equation}
with $a=2/\sinh 2\rho$. Here we note that the  left-moving primary
$h^L_{\mu\nu}$ is found when $\mu l=1$.

\section{TMG at the chiral point}
We start by noting that at the chiral point, one has
$\psi^M_{\mu\nu}=\psi^L_{\mu\nu}$ coupled to (2,0)-operator with
 spin $s=2$.
 In this section, we study a new mode from
the massive graviton defined by~\cite{GJ}
\begin{equation} \label{new-L}
\psi^{new}_{\mu\nu}=\partial_{\mu l}\psi^M_{\mu\nu}|_{\mu
l=1}=y(\tau,\rho)\psi^L_{\mu\nu}
\end{equation}
with
\begin{equation}
y(\tau,\rho)=-i \tau -\ln [\cosh \rho].
\end{equation}
We note that the partial derivative of $\partial_{\mu l}$ is the
logarithmic operator to generate the quasi-primary field from the
primary field $\psi^M_{\mu\nu}$~\cite{Lewis1} and it commutes
other operators. Using $L_0y=\bar{L}_0y=1/2$ and
$L_1y=\bar{L}_1y=0$, it is easily shown that for  a non-chiral
field $\psi^{new}_{\mu\nu}$, one has
\begin{equation}
L_0\psi^{new}_{\mu\nu}=2\psi^{new}_{\mu\nu}+\frac{1}{2}\psi^{L}_{\mu\nu},~~
\bar{L}_0\psi^{new}_{\mu\nu}=\frac{1}{2}\psi^{L}_{\mu\nu},~~~
L_1\psi^{new}_{\mu\nu}=\bar{L}_1\psi^{new}_{\mu\nu}=0.
\end{equation}
while for a chiral field $\psi^{L}_{\mu\nu}$,  these are
\begin{equation}
L_0\psi^{L}_{\mu\nu}=2\psi^{L}_{\mu\nu},~~
\bar{L}_0\psi^{L}_{\mu\nu}=0,~~~
L_1\psi^{L}_{\mu\nu}=\bar{L}_1\psi^{L}_{\mu\nu}=0.
\end{equation}
This shows that $\psi^{new}_{\mu\nu}$ is not an eigenstate of
$L_0$ and $\bar{L}_0$ and thus it is impossible to decompose it as
a linear combination of eigenstates to $L_0$ and $\bar{L}_0$.
Actually $\psi^{new}_{\mu\nu}$ is a logarithmic partner of
$\psi^{L}_{\mu\nu}$. We obtain the equations for $\psi^{new}$
\begin{equation}
\Big( {\cal D}^L h^{new} \Big)_{\mu\nu}\not=0,~~\Big( {\cal D}^L
{\cal D}^L h^{new} \Big)_{\mu\nu}=0,~~\Big( {\cal D}^R {\cal D}^L
{\cal D}^L h^{new} \Big)_{\mu\nu}=0.
\end{equation}
Even though $\psi^{new}$ solves the classical equation, it does
not provide the conventional equation.  We derive the equation of
motion from $( {\cal D}^R {\cal D}^L h^{new})_{\mu\nu}=
-\psi^{L}_{\mu\nu}/l^2$ as a coupled equation
\begin{equation}
\left ( \nabla^2 + \frac{2}{l^2} \right )\psi^{new}_{\mu\nu} -
\frac{2}{l^2}\psi^{L}_{\mu\nu} =0. \label{new-equation}
\end{equation}
Combining Eq.(\ref{new-equation}) with Eq.(\ref{sec-equation})
leads to the fourth-order equation for $\psi^{new}_{\mu\nu}$
\begin{equation}
\left ( \nabla^2 + \frac{2}{l^2} \right )^2 \psi^{new}_{\mu\nu}
=0, \label{new4-eq}
\end{equation}
which is basically different from the second-order equations for
$\psi^L_{\mu\nu}$. This fourth-order equation may induce the
unitarity problem.

Unfortunately, this field and its descendants diverge linearly in
$y(\tau,\rho)$ near the boundary at $\rho=\infty$~\cite{Stro}.
 Hence it is problematic to argue that $\psi^{new}$ is a physical
mode but not pure gauge at the chiral point.

On the other hand, it was suggested that a field
$X_{\mu\nu}=(\bar{L}_{-1}\psi^{new})_{\mu\nu}$ coupled to
(2,1)-operator may be a primary field with the correct asymptotic
behavior at the chiral point~\cite{GKP}, even though  $X_{\mu\nu}$
failed to be a truly primary field because of $L_1X_{\mu\nu}=0$
and $\bar{L}_1X_{\mu\nu}=\psi^L_{\mu\nu}$. It was argued that if
one defines physical states modulo locally pure-gauge states,
$X_{\mu\nu}$ would be a primary field.

 However, in the next section, we prove
that $\psi^{new}_{\mu\nu}$ is not a physical field and
$X_{\mu\nu}$ could not be a primary field by using the logarithmic
conformal field theory (LCFT).

\section{LCFT and Singleton coupled operator with $s=0$}
It was shown that $\psi^{new}_{\mu\nu}$ is a logarithmic partner
of $\psi^L_{\mu\nu}$, implying the LCFT~\cite{GJ,Sachs}.
 A LCFT differs from an ordinary CFT in that the Virasoro
generator $L_0$ is not diagonalizable~\cite{LCFT}. In addition to
the primary and descendant fields, it includes pairs of operators
which form the Jordan cell structure for $L_0$ with highest
weights $(h,\bar{h})$
\begin{eqnarray} \label{JCS}
L_0|C\rangle&=&h |C\rangle,~~L_0 |D\rangle=h|C\rangle
|D\rangle+|C\rangle, ~~~~L_n|C\rangle=L_n|D\rangle=0,~~n\ge1
\nonumber \\
\bar{L}_0|C\rangle&=& \bar{h} |C\rangle,~~\bar{L}_0
|D\rangle=\bar{h}|C\rangle |D\rangle+|C\rangle,
~~~~\bar{L}_n|C\rangle=\bar{L}_n|D\rangle=0,~~n\ge1
\end{eqnarray}
which means that the primary field $C$ and the quasi-primary field
$D$ are in reducible, but indecomposable representation of the
Virasoro algebra. Kogan proposed that a dipole ghost pair ($A,B$)
can represent a singleton, which induces the 2-point function for
a pair of logarithmic operators ($D,C$) with
$h=\bar{h}$~\cite{Kogan}. Here $C$ is an operator with (1,1). Also
he has shown that this is the origin of logarithmic singularities
in the 4-point functions. This logarithmic pair  has the 2-point
functions~\cite{GKA}
\begin{eqnarray}
\langle C({\bf x}) C({\bf y}) \rangle &=& 0,
\nonumber \\
\langle C({\bf x}) D({\bf y}) \rangle &=&
      { c \over {|{\bf x}-{\bf y}|^{2 \Delta}}},
\nonumber \\
\langle D({\bf x}) D({\bf y}) \rangle &=&
      { c \over {|{\bf x}-{\bf y}|^{2 \Delta}}} \left [d
              -2c \ln|{\bf x}-{\bf y}|  \right ].
\label{2-point}
\end{eqnarray}
Here $\Delta=h+\bar{h}$ is a degenerate dimension  of $C$ and $D$.
The coefficient $c$ is determined by the normalization of $C$ and
$D$. However, $d$ is arbitrary and thus it can be set to any
value, using the symmetry of the theory under $D \to D+\lambda C$
which leaves Eq.(\ref{JCS}) unchanged.

In the AdS$_3$/CFT$_2$ correspondence, there exists a puzzle of
missing states between CFT$_2$ and supergravity~\cite{Vafa}. The
gauge bosons appear in the resolution of this puzzle. These are
chiral primaries~\cite{MS}. But on the supergravity side, these
are absent and thus may be considered as unphysical singletons on
AdS$_3$~\cite{Boer}.  The authors in~\cite{KLM} found that these
gauge bosons coupled to (2,0) and (0,2) operators on the boundary
receive logarithmic corrections from an AdS$_3$ scattering
calculation. The clearest evidence for the existence of logarithmic
operators in AdS$_3$/CFT$_2$ correspondence comes from
calculations of the greybody factors in AdS$_3$ spacetimes. Since
the greybody factors are related to the 2-point functions in CFT,
logarithms found here are a clear indication that we have
logarithmic operators on the boundary. In this sense, it was
important to test the relationship between gauge boson and
singleton by calculating greybody factors~\cite{MLee}.

Correlation functions of operators ${\cal O}_i({\bf x})$ in the
CFT$_2$ on the boundary of AdS$_3$ spacetimes which corresponds to
fields $\Phi_i$ in the bulk could be calculated from the bulk
action $I$ using the relation
\begin{equation}
\langle e^{\Sigma_i\int d^2x \Phi_{b,i}({\bf x}){\cal O}_i({\bf
x})}\rangle_{CFT}=e^{-I[\{\Phi_i\}]}|_{\Phi_i=\Phi_{b,i}}.
\end{equation}
We are now in a position to introduce  the  action for a dipole
ghost pair ($A,B$) on the AdS$_3$ spacetimes~\cite{GKA,Kogan}
\begin{equation}
I_{\rm DG} =\frac{1}{16\pi G_3} \int d^3 x \sqrt{-g} \left [
\partial_\mu A
\partial^\mu B
   - m^2 A B - {1 \over 2} B^2 \right ]
\label{dipole-action}
\end{equation}
with mass $m$.  It is not clear if this action comes from
supergravity theories. Rather, this  takes a similar form of the
Nakanishi-Lautrup formalism in the gauge theory~\cite{Kugo}. In
detail, Eq.(\ref{dipole-action}) with $m^2 = 0$ and $A_\mu =
\partial_\mu A$ leads to a gauge-fixing term as
\begin{equation}
I_{\rm GF} = - \int d^3 x \sqrt{g} \left [ B \partial_\mu A^\mu +
{ \alpha \over 2} B^2 \right ] \end{equation}
 with $\alpha=1$.
Here $B$ is the Nakanishi-Lautrup field, while $A$ corresponds to
$\sigma$-field  which leads to the negative-norm state. Thus $A$
and $B$ form the zero-norm states.

Their equations of motion are given by
\begin{equation} \label{AB-eq}
\left ( \nabla^2 + m^2 \right ) A + B =0, ~~ \left ( \nabla^2 +
m^2 \right ) B =0 \end{equation}
which leads to the fourth-order
equation for $A$
\begin{equation}
\left ( \nabla^2 + m^2 \right )^2 A  =0. \label{4th-equation}
\end{equation}
This is the same equation  (\ref{new4-eq}) for
$\psi^{new}_{\mu\nu}$~\cite{GJ} upon choosing $\psi^L_{\mu\nu}
\sim -B$. A pair of  dipole ghost fields ($A,B$) coupled to
$(D,C)$-operators on the boundary could represent  a pair of
$(\psi^{new}_{\mu\nu},\psi^L_{\mu\nu})$. Hence we may have a
correspondence
\begin{equation}
B \leftrightarrow \psi^{L},~~A \leftrightarrow \psi^{new}.
\end{equation}
Here we observe a slight difference that  a scalar $B$ is coupled
to a primary operator  $C$ with (1,1), whereas a tensor field
$\psi^L_{\mu\nu}$ is coupled to a primary operator with (2,0). On
the other hand, a scalar $A$ is coupled to a quasi-primary
operator $D$, while a tensor $\psi^{new}_{\mu\nu}$ is coupled to a
quasi-primary operator.

In addition, we  may derive equations for $A$ and $B$ using the
operator method. For a primary field  $C$, from Eq.(\ref{JCS}), we
have
\begin{equation}
C \sim \frac{e^{-ihu-i\bar{h}v}}{(\cosh \rho)^{h+\bar{h}}}
\end{equation}
From the condition for $D$ in Eq.(\ref{JCS}), we have
\begin{equation}
D=[u+v+f(\rho)]C,
\end{equation}
where $f(\rho)$ is determined by considering the singleton
condition of  $L_1|D\rangle=\bar{L}_1|D\rangle=0$ as
\begin{equation}
f(\rho)=-2i \ln[\cosh\rho]+\delta.
\end{equation}
Here $\delta$ is an arbitrary constant, which can set to any value
using the freedom to shift $D$ by an amount proportional to
$C$($D\to D+\lambda C$).
 Finally, we obtain
\begin{equation}
D=[-i\tau -\ln[\cosh\rho]+\delta'](-2iC),
\end{equation}
which is the same relation  as in Eq.(\ref{new-L}). Evaluating the
second Casimirs  gives the equations of motion for the bulk fields
$\tilde{D}$ and $\tilde{C}$ in AdS$_3$ spacetimes which are
related  to $D$ and $C$ as
\begin{equation}
\Big(\bar{\nabla}^2 +
m^2\Big)\tilde{D}-4(\Delta-1)\tilde{C}=0,~\Big(\bar{\nabla}^2 +
m^2\Big)\tilde{C}=0
\end{equation}
with the mass $m^2l^2
=2h(h-1)+2\bar{h}(\bar{h}-1)=\Delta(\Delta-2)$. Comparing the
above with Eq.(\ref{AB-eq}), one finds a relation of $\tilde{D}
\sim A$ and $\tilde{C} \sim -B$.  Here we have $m^2=0$ because of
$h= \bar{h}=1(\Delta=2)$. This is clear since the massless dipole
pair  of ($A,B$) corresponds to the massless graviton pair of
$(\psi^{new}_{\mu\nu},\psi^{L}_{\mu\nu})$.

 We are now in a position to show that
$X_{\mu\nu}=\bar{L}_{-1}\psi^{new}_{\mu\nu}$~\cite{GKP} could not
be a primary field at the chiral point. Let us assume that there
exist two primary fields $C$ and $C'$. In the earlier situation in
Eq.(\ref{JCS}) where two $C$ and $D$ became degenerate, while in
this case  we have two fields whose dimension differs by an
integer $N$ as~\cite{Lewis2}
\begin{equation}
L_n|C\rangle =0, n\ge 1~{\rm and}~~(L_1)^N|D\rangle=\beta
|C'\rangle, ~~L_n|D\rangle=0,n\ge 2,
\end{equation}
where $C'$ is another primary field with conformal weights
$(h-N,\bar{h})$ and $\beta$ is a non-zero constant. In this case,
$C$ with $(h,\bar{h})$ is  not really a primary field but rather a
descendent of $C'$: $|C\rangle =\sigma_{-N}|C'\rangle$, where
$\sigma_{-N}$ is some combination of the Virasoro generators of
dimension $N$ with $[L_n,\sigma_{-N}]=0,~n\ge1$. This implies why
the 2-point function is still zero. Explicitly, a logarithmic pair
$C$ and $D$ still have the same 2-point functions as in
Eq.(\ref{2-point}). However, $C'$ is just an ordinary primary
field with the usual 2-point function
\begin{equation}
\langle C'(u,v) C'(0,0)\rangle \propto
\frac{1}{u^{2(h-N)}v^{2\bar{h}}}.
\end{equation}
 If $\beta
\not=0$, we have (1,1) for $C$ and then (0,1) for $C'$ when $N=1$.
However, this does not belong to $X_{\mu\nu}$ coupled to
(2,1)-operator. Furthermore, we could not develop another primary
field $C'$  because the condition of $L_1\psi^{new}_{\mu\nu}=0$
for $N=1$ implies $\beta=0$. There is no room to accommodate
another primary field  at the chiral point except a pair of
$(\psi^L,\psi^R)$.

\section{Discussions}
We investigate whether or not  the new tensor fields of
$\psi^{new}_{\mu\nu}$ and $X_{\mu\nu}$  become  propagating
physical fields at the chiral point. This work is important
because  an urgent work in the chiral gravity is to find
physically propagating  degrees of freedom on the AdS$_3$
spacetime background using the AdS$_3$/CFT$_2$ correspondence. We
show that there is no physically propagating degrees of freedom at
the chiral point by using the logarithmic conformal field theory.
Explicitly, it is found that ($\psi^{new},\psi^L$) form a dipole
ghost pair (unphysical fields) as well as $X$ is not a primary
field. Hence our result supports that the original work of chiral
gravity~\cite{LSS1} is correct.

On the other hand,  we discuss the unitarity problem related to
the logarithmic operators.  This is clearly understood from the
bulk side.  In general, the fourth-order equations for
$\psi^{new}$ and $A$ may induce the unitarity problem in
calculating their on-shell amplitudes.  It is noted that these
logarithmic terms originate from the unphysical dipole ghost
fields ($A,B$). As was shown in \cite{Kugo}, this pair ($A,B$) is
turned into the zero-norm state by the Goldstone dipole mechanism
in Minkowski spacetime. We suggest that the boundary logarithmic
terms is related to  the negative-norm state of $A$. In order to
remove the negative-norm state, we impose the subsidiary condition
as $B^+(x) \vert 0 \rangle_{\rm phys}=0$. Then the physical
space($\vert 0 \rangle_{\rm phys}$) will not include any
$A$-particle state. This corresponds to the dipole mechanism to
cancel the negative-norm state. Similarly, we expect that  this prescription may work for the
 CFT$_2$ on the boundary at infinity.

Fortunately, since the new field $\psi^{new}_{\mu\nu}$ is not a
physical field, we do not consider this unitarity problem
seriously. As a byproduct of introducing $\psi^{new}_{\mu\nu}$, we
show that the chiral gravity is unitary, leaving the right-moving
graviton because a pair of $(\psi^{new}_{\mu\nu},\psi^L_{\mu\nu})$
forms zero-norm state as a dipole ghost pair.

\section*{Acknowledgement}
The author thanks Myungseok Yoon for helpful discussions. This was
in part supported by the Korea Research Foundation
(KRF-2006-311-C00249) funded by the Korea Government (MOEHRD) and
the SRC Program of the KOSEF through the Center for Quantum
Spacetime (CQUeST) of Sogang University with grant number
R11-2005-021.

\end{document}